\documentstyle[seceq,epsfig]{ptptex}

\newcommand{\lab}[1]{\label{#1}}

\newcommand{\fslash}[1]{#1
  \hspace{-5.5pt}\raisebox{-0.2pt}{\slash}\hspace{2pt}}

\title{Dynamical screening in hot systems away from (chemical) equilibrium
\footnote{presented by M.D. at the Japanese Workshop on Thermal Quantum Field 
Theories and their Applications, Kyoto, Japan, 25.-27.~August 1999} }

\author{R. {\sc Baier$^1$},  M. {\sc Dirks$^1$$^2$}, K. {\sc Redlich$^3$}  }
\inst{
$^1$
Fakult\"{a}t f\"{u}r Physik, Universit\"{a}t Bielefeld,
D-33501 Bielefeld, Germany \\
$^2$Faculty of Science, Osaka City University, Sumiyoshi-Ku, Osaka 558, Japan\\
$^3$ Institute for Theoretical Physics, University of Wroclaw,
PL-50204 Wroclaw, Poland \\
}

\abst{Within the Closed Time Path Formalism of Thermal Field Theory we
calculate the hard photon emission rate as well as the collisional energy-loss
rate for a quark-gluon plasma away from chemical equilibrium. Mass singularities are shown to be dynamically screened within HTL-resummed perturbation theory
also away from equilibrium. Additional (pinch) singularities are absent and
well defined results are obtained.}

\begin{document}

\maketitle

\section{Introduction}
At present, major efforts are being made to experimentally probe a
new deconfinement state of strongly interacting matter 
(Quark-Gluon Plasma, QGP), and much related research is being done in
order to establish a sufficient theoretical understanding of the systems
under investigation here. As a central result of research done to date, the
hard-thermal-loop (HTL) resummation scheme \cite{htl1,htl2,htl3}
allows to take collective
effects into account within an improved perturbative approach. For systems
in thermal equilibrium, based
on HTL-effective perturbation theory, predictions for potentially 
observable emission rates can be obtained, with medium effects providing the 
required (dynamical) screening of singularities. 
 
More recently, need for a better theoretical understanding of non-equilibrium
effects has become apparent, since systems actually under investigation
in a heavy-ion collision are likely to stay away from equilibrium for
important periods. In this work, we attempt to study the role of collective
effects in systems away from equilibrium and to extend the HTL-resummation
prescription appropriately. We will study hard thermal photon
production \cite{kapusta:photon,baier:photon,shuryak:nephoton,strickland:nephoton,kaempfer:nephoton,thoma:nephoton,muller:nephoton}
as well as the collisional energy loss \cite{bjorken:engl,braaten:engl1,braaten:engl2}, which both are known
to be sensitive to soft scale physics. The following is based on 
\cite{baier:nephoton,baier:neegloss}, where further details can be found. 

\section{The approximation scheme -  chemical non-equilibrium}
The scenario of chemical non-equilibrium 
\cite{shuryak:hotglue,biro:therma,kapusta:cascade,wong:therma}
has been argued to be a 
valid simplification of the actual, complicated pre-equilibrium 
dynamics in a heavy-ion collision. It is inspired from an analysis of the
processes leading to equilibration eventually, from which the effect of
elastic collisions is more pronounced than the effect of inelastic
processes. Consequently, while local thermal equilibrium might be
quickly established, chemical equilibrium among the different particle 
species and their respective number-densities will be delayed. In particular
the number density of quarks is expected to be low while gluons will
reach their equilibrium density more quickly, as has been formulated in the
hot-glue scenario \cite{shuryak:hotglue}.

In this work we follow the analysis, e.g., of \citen{biro:therma}, 
parameterizing chemical non-equilibrium in terms of fugacity factors $\lambda$
multiplying the distribution functions 
$\tilde n(X,p) = \lambda_q(X) n_F$ for quarks and $n(X,p)=\lambda_g(X) n_B $ 
for gluons\cite{thoma:nephoton,kaempfer:nedilepton}. 
Here $n_F, n_B$ are the equilibrium Fermi- and Bose-distributions
respectively, and the (space-) time variable $X$ refers to the scale of 
chemical equilibration. We do not attempt to analyze the process of 
equilibration but assume the evolution of $\lambda_{q,g}$ as input, e.g., from 
\citen{biro:therma}. As a function of these, we attempt to determine photon
production and energy-loss rates. Based on the Closed-Time-Path formalism of 
Thermal Field Theory \cite{landsman:rev,chou:noneq} and within the simplified scenario of chemical
non-equilibrium, we consider only lowest (second) order initial 
correlations and 
the scale of equilibration can be argued to be $X\sim 1/g^4 T$. Scales
fast with respect to $X$, in particular both hard and soft scales, can than
be transformed to momentum space (Wigner transformation). To lowest order of
an expansion in the gradients of equilibration 
\cite{chou:noneq,heinz:noneq,calzetta:noneq,henning:noneq}, 
the resulting perturbation theory  is very 
similar to the equilibrium formalism with bare propagators for 
scalars and fermions respectively
\begin{eqnarray}
 & &iD_{21} = 2\pi \varepsilon(p_0) \delta(p^2) (1+n(X,p_0)) ,\quad
 iS_{21}(X,p) = \fslash{p} 2\pi\delta(p^2) \varepsilon(p_0) 
 (1-\tilde n(X,p_0)), \lab{eq:propfree}\nonumber
\end{eqnarray}
now depending on the modified distribution functions. Physical rates
calculated in this way are expected to depend on $\lambda_q,\lambda_g$ via
$\tilde n(X,p)$ and $n(X,p)$. However, when investigating quantities sensitive
to soft-scale physics it will be necessary to extend the HTL-resummation 
scheme to the present situation away from equilibrium. 
Doing so the required  partial resummation of 
self-energy corrections is known to entail contributions not present in the
equilibrium analysis \cite{altherr:pinch1,altherr:pinch2}, which should be 
taken into account.

\section{Soft fermion exchange in hard real photon production}
To lowest order in perturbation theory, real photon production from the
QGP arises from Compton and annihilation processes. The respective production
rates can be calculated from the absorptive part of the
2-loop selfenergy diagram shown in Fig.~1a. At fixed lowest order, a divergent 
result is known to arise from a mass-singularity induced by the exchange of a
massless quark in the process. However, in case of an equilibrium system,
with the dispersion of soft quarks being modified within the HTL-framework, 
the singularity is found to be screened by Landau-damping effects and a 
well defined result is obtained  \cite{kapusta:photon,baier:photon}.
 
\begin{figure}
\mbox{(a)
\epsfig{file=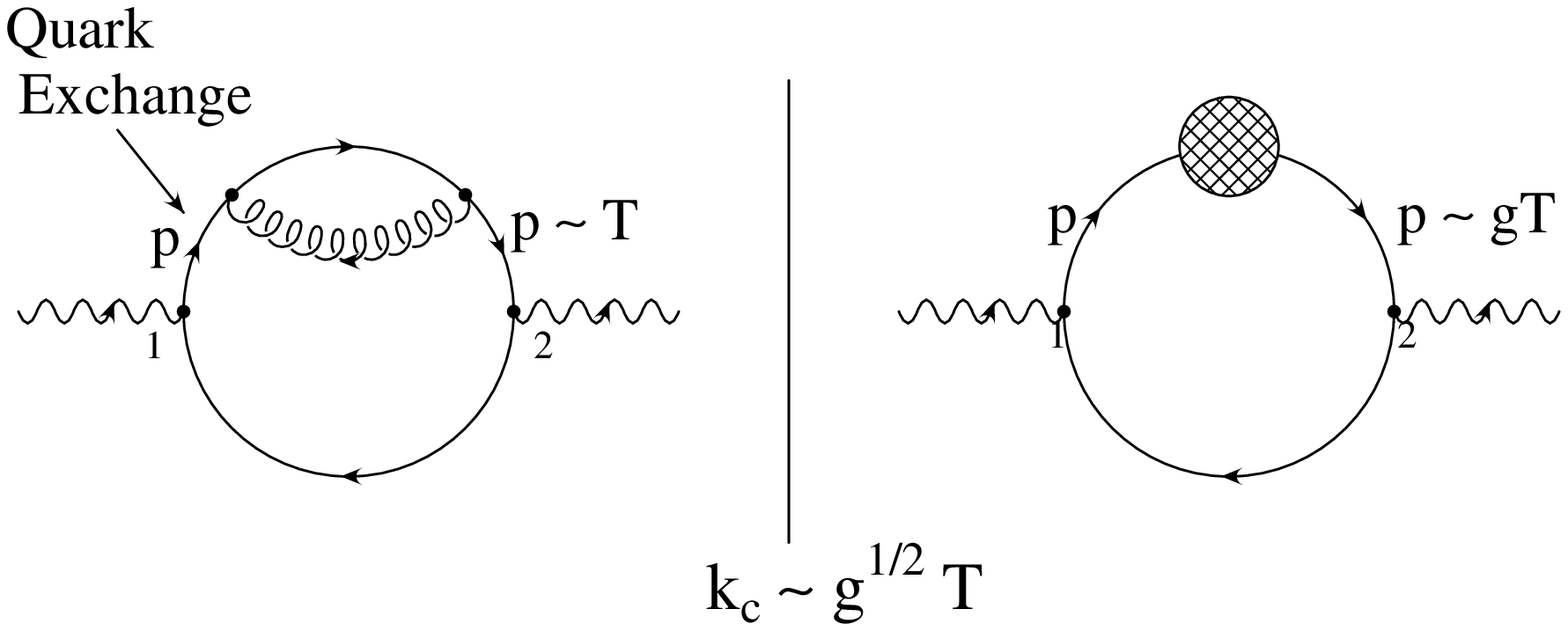, width=7cm}\quad
(b)\epsfig{file=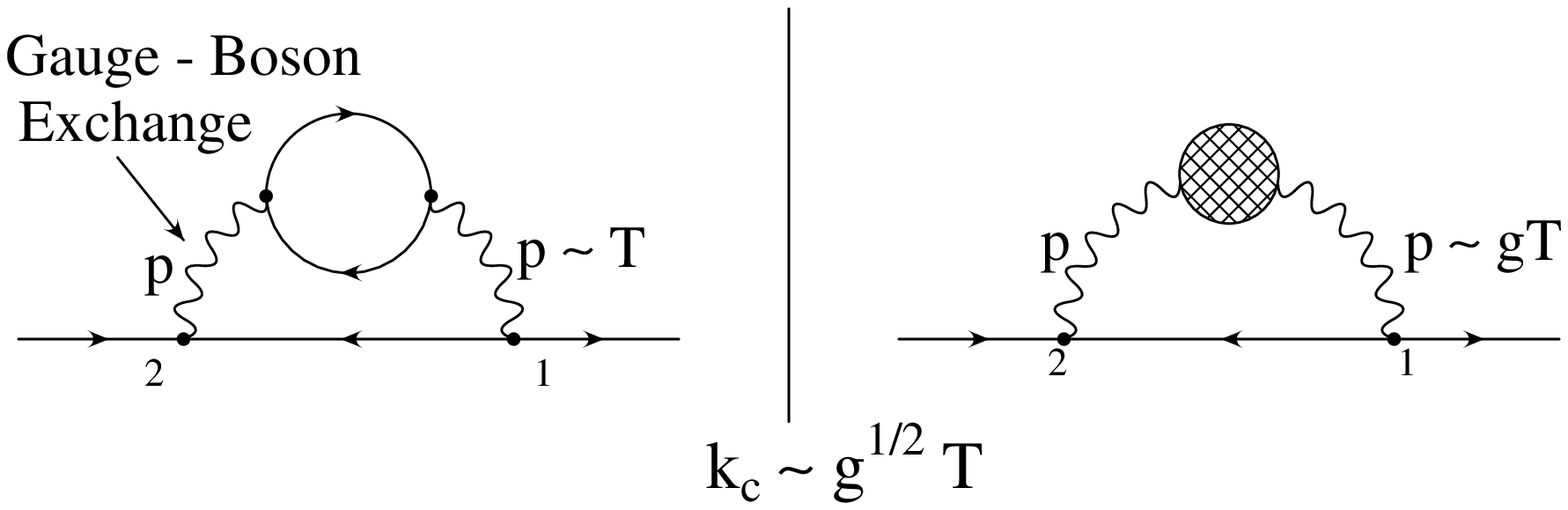, width=7cm}}
\caption{\label{fig:photon}Diagrams contributing to hard photon production 
(a) and collisional energy-loss rates (b). The lefthand side of each
display shows fixed order of perturbation theory for hard exchanged momentum,
while the right-hand side indicates HTL-resummation for soft exchanged 
momentum. The parameter $k_c$ separating
hard and soft regions drops out in the sum of hard and soft partial rates.
  \protect\cite{braaten:screening}  }
\end{figure}

Away from equilibrium, the resummed propagator for soft quark exchange
can be written
\begin{eqnarray}
 & & S_{12}^\star (p)    = - \tilde n(X,p_0)
 (\frac{1}{\fslash p -\Sigma_R + i\varepsilon p_0} -
        \frac{1}{\fslash p -\Sigma_R^\star - i\varepsilon p_0})  \nonumber \\
 & &-  \frac{1}{\fslash p - \Sigma_R + i\varepsilon p_0 }
  \left[ (1-\tilde n(X,p_0)) \Sigma_{12} 
+ \tilde n(X,p_0) \Sigma_{21} 
  \right] \frac{1}{\fslash p - \Sigma_R^\star -i\varepsilon p_0} . \lab{eq:propf}
\end{eqnarray}
Here, the first term on the right hand side has a structure parallel to
the equilibrium expression, while the second term vanishes if detailed 
balance holds and contributes only away from equilibrium. There are two points
to be made. First, non-equilibrium modifications to the dispersion, as
defined by the retarded propagator in the first term of Eq.~(\ref{eq:propf}),
amount to a modification of the soft-scale parameter $m_q$ only:
\begin{eqnarray}
  \tilde m_q^2 &=& \frac{g^2}{2\pi^2} C_F \int_0^\infty E dE (n(X,E) + \tilde n(X,E))= \frac{g^2 T^2}{12} C_F \left(\lambda_g + \frac{\lambda_q}{2}\right) .
 \lab{eq:newmq}
\end{eqnarray}
The reason for this is that HTL-contributions to the self-energy, just as 
in the equilibrium case, come proportional to the integral displayed  
in the definition of $\tilde m_q$, Eq.~(\ref{eq:newmq}). 
Next, in order to discuss the second term on the right-hand side of 
Eq.~(\ref{eq:propf}) we propose to rewrite it according to
\begin{eqnarray}
& &(1-\tilde n(X,p_0)) \Sigma_{12} + \tilde n(X,p_0) \Sigma_{21}  = 
  (1-2\tilde n(X,p_0)) \Sigma^- + \Sigma^+ .
\end{eqnarray}
Here,  $\Sigma^- = \frac{1}{2}(\Sigma_{12} - \Sigma_{21} ) =
iIm \Sigma_R$ and $\Sigma^+ = \frac{1}{2} \left(\Sigma_{12} + \Sigma_{21}\right)$
have been introduced. 
In the fermionic case, the contribution from $\Sigma^+$ is found to 
be small compared to the HTL-contribution in $\Sigma^-$. Explicitly we
find
\begin{eqnarray}
& &\Sigma^+ (X,p) \sim p_0 R\, \Sigma^-(X,p) \ll \Sigma^- ,
\quad R = \frac{ \int_0^\infty dk~  k~  \frac{\partial}{\partial k} 
  \tilde n(k) (1+2n( k)) }{ \int_0^\infty dk~  k~ [n(k) + \tilde n(k)]} 
 \sim \frac{1}{T} ,
\end{eqnarray}
so that for $p_0\sim gT$, $\Sigma^+\sim g^3T^2$ results.
Consequently, within the HTL-scheme, the contribution from $\Sigma^+$
to the HTL-resummed propagator can be neglected. Doing so and using
furthermore
\begin{equation}\lab{eq:rearf}
 2\Sigma^- = \Sigma_R -\Sigma_A = 
  {S_A^\star}^{-1} - {S_R^\star}^{-1}, 
\end{equation}
the effective propagator can be simplified into
\begin{eqnarray}
\left.  S_{12}^\star \right|_{HTL} &\simeq& -\tilde n(X,p_0) \left(  
 S_R^\star -  S_A^\star\right) 
  - (\frac{1}{2}-\tilde n(X,p_0)) S_R^\star [2\Sigma^-] S_A^\star
 \nonumber \\
 &=& - \frac{1}{2} \left(S_R^\star - 
 S_A^\star \right) \label{eq:propfeff}.
\end{eqnarray}
In this form, $S_{12}^\star$ has the same structure as in equilibrium so 
that the remaining phase space integrations can be performed as explained
in the respective literature \cite{kapusta:photon,baier:photon} taking
only the modification of the dispersion, Eq.~(\ref{eq:newmq}), into account.
The hard photon production rate from a system away from equilibrium 
is obtained as 
\begin{eqnarray}
  E_\gamma \frac{dR}{d^3q} &=& e_q^2 \frac{\alpha\alpha_s}{2\pi^2}
 {\lambda_q} T^2
   e^{-E_\gamma/T} \left[ \frac{2}{3}{(\lambda_g + \frac{\lambda_q}{2})}
  \ln \left(\frac{2E_\gamma T}{{\tilde m_q^2(\lambda_q,\lambda_g)}} 
\right)\right.
 \nonumber \\ 
 & & \left. \hspace{5cm} + \frac{4}{\pi^2} C(E_\gamma ,
  T, \lambda_q, \lambda_g) \right] ,  \lab{eq:resph}
\end{eqnarray}
where the finite contribution $C(E_\gamma ,
  T, \lambda_q, \lambda_g)$ has been given explicitly in 
\citen{baier:nephoton}.
The important observation in Eq.~(\ref{eq:resph}) is that the 
logarithmic divergence turns out to be screened on the scale of the 
parameter $\tilde m_q(\lambda_q,\lambda_g)$.
The result, Eq.~(\ref{eq:resph}), is independent of the parameter
$k_c$ upon summation of soft and hard partial rates. The noteworthy point
here is that no additional (pinch) singularities 
\cite{altherr:pinch1,altherr:pinch2} 
arise from the second term in the propagator, Eq.~(\ref{eq:propf}),  when calculating to one-loop order. This is because 
phase space for real photon production is restricted to space-like 
momentum exchange, and the pinch-singular region on the mass-shell does not
contribute in this case. The propagator for the line of hard quark exchange
in Fig.~1a, as given by the one-loop approximation to Eq.~(\ref{eq:propf}),
can than be written as
\begin{equation}
\left. \delta S_{12}(p) \right|_{p^2 \le -k_c^2} 
   \stackrel{\wedge}{=}
   -\frac{1}{(p^2)^2} \fslash p \Sigma_{12} \fslash p, 
\end{equation}
which, similar to Eq.~(\ref{eq:propfeff}), does not depend explicitly on 
the distribution $\tilde n(X,p)$. The remaining mass-singularity turns out
to be dynamically screened in Eq.~(\ref{eq:resph}).

\section{Exchange of a soft boson in the collisional energy loss}
Next we consider the collisional energy-loss for a charged 
particle going through a QED medium assumed to be away from equilibrium. 
We consider a heavy external 
probe with mass $M\gg T$\cite{braaten:engl1,braaten:engl2}. To lowest order
the energy-loss arises from elastic collisions with particles in the
medium and can be obtained from the (21)-component of the 
selfenergy of the projectile as shown in Fig.~1b: 
\begin{eqnarray}
  - \frac{dE}{dx} &=& - \frac{1}{4E} Tr\left[ (\fslash q + M) 
 i\Sigma'_{21}(X,q)\right] 
 \nonumber \\
 &=& \frac{e^2}{4E} \int \frac{d^3 \vec{p}}{(2\pi)^4}
   \int\frac{dp_0 p_0}{v} Tr[(\fslash q + M) \gamma_\mu iS_{12}^{T=0} (p-q) 
 \gamma_\nu]
 iD_{21}^{\mu\nu}(X,p).
\lab{eq:egl1}
\end{eqnarray}
Here, an additional power of $p_0$, the
energy exchanged, is present in the integrand, which, 
for an equilibrium systems,
makes the energy-loss finite after resummation of HTLs, in contrast with the
fermion damping rate.

In this case, in contrast with the exchange of a soft fermion as considered
in the previous section, we investigate the exchange of a soft (gauge) 
boson. The longitudinal and transverse components of the 
corresponding resummed propagator can be expressed as 
\cite{thoma:noneq1}
\begin{eqnarray}
 D_{21}^{\star \scriptscriptstyle{L/T}} &=& (1+n(X,p_0)) 
 (D_R^{\star \scriptscriptstyle{L/T}} - D_A^{\star \scriptscriptstyle{L/T}} )
  \nonumber \\
 & & \hspace*{-0.8cm} + D_R^{\star \scriptscriptstyle{L/T}} \left[ n(X,p_0) 
 \Pi_{21}^{\scriptscriptstyle{L/T}} - (1+n(X,p_0)) \Pi_{12}^{\scriptscriptstyle{L/T}}\right]
 D_A^{\star \scriptscriptstyle{L/T}}.
\label{eq:propb}
\end{eqnarray}
The propagator $D_{21}$, Eq.~(\ref{eq:propb}), can be analyzed along the same
lines as the fermion propagator, Eq.~(\ref{eq:propf}). 
Namely, with regard to the dispersion the only non-equilibrium modification
concerns the plasma frequency, which is redefined according to
\begin{equation}\lab{eq:newmb}
m_\gamma^2 = \frac{e^2 T^2}{9} \quad\to\quad
 \tilde m_\gamma^2 = \frac{4e^2}{3\pi^2} \int_0^\infty dk k \tilde n(X,k) = 
  \lambda_f m_\gamma^2 .
\end{equation}
However, when investigating the second term in $D_{21}$, Eq.~(\ref{eq:propb}),
the contribution of $\Pi^+= \frac{1}{2}(\Pi_{12} + \Pi_{21})$ turns out
to be larger than that of 
$\Pi^- = \frac{1}{2}(\Pi_{12} - \Pi_{21}) = iIm \Pi_R$\cite{thoma:noneq1}
\begin{eqnarray}
& &-\Pi^+_{\scriptscriptstyle{L/T}} = \frac{1}{2p_0}~ R~
\Pi^-_{\scriptscriptstyle{L/T}} , \quad
R = \frac{\int_0^\infty dk k^2 \tilde n(k) (1-\tilde n(k))}
 {\int_0^\infty dk k\tilde n(k)} \sim T , 
\end{eqnarray}
and $\Pi^+$ may not be neglected in general. However, 
in the energy-loss the contribution of $\Pi^+$ integrates to zero given 
its odd symmetry under $p_0 \to -p_0$:
\begin{equation}
\int_{-vp}^{vp} dp_0 { p_0} \Pi^+(X,p_0) = 0 .
\end{equation}
Therefore, there is no contribution from  $\Pi^+$ to $dE/dx$, and the
propagator, Eq.~(\ref{eq:propb}), effectively simplifies to  
\begin{equation}
\left. D_{21}^{\star \scriptscriptstyle{L/T}} ~\right|_{dE/dx} 
 \stackrel{\wedge}{=} \frac{1}{2}
   (D_R^{\star \scriptscriptstyle{L/T}} - D_A^{\star \scriptscriptstyle{L/T}})
 \qquad \mbox{in}\quad dE/dx. 
\end{equation}
Again this form is similar to the equilibrium expression, and the phase-space
integration may be performed as in \citen{braaten:engl1}. As for real
photon production, pinch singularities are absent since 
only space-like momentum exchange is involved, and the remaining mass 
singularity is dynamically screened, providing a well defined result. 

When generalizing to the case of a heavy quark propagating through a 
QGP \cite{braaten:engl2} one obtains for $E\ll M^2/T$, 
\begin{eqnarray}\lab{eq:reseg}
-\frac{dE}{dx} &=& \frac{ g^2 \tilde m_g^2}{2\pi v} 
 \left(1-\frac{1-v^2}{2v}
 \ln \frac{1+v}{1-v}\right) \left[ \ln \frac{ET}{3M{ 
 \tilde m_g}} - \frac{\ln 2}{(1+\lambda_qN_f/6\lambda_g)} + A(v)\right],
\label{eq:reseng}
\end{eqnarray}
and for $E\gg M^2/T$,
\begin{equation}
  - \frac{dE}{dx}  = \frac{g^2 {\tilde m}^2_g}{2\pi} 
\ln \left[ 0.920 \frac{\sqrt{ET}}{ \tilde m_g} \,
 2^{\lambda_q N_f/(12 \lambda_g + 2 \lambda_q N_f)} 
                                \right].
\end{equation}
Here, $g$ is the QCD coupling constant, 
$\tilde m_g=g^2T^2(\lambda_g+\lambda_qN_f/6)/3$ is the appropriate generalization of
$\tilde m_f$, Eq.~(\ref{eq:newmb}), now depending on both the quark and the
gluon fugacities and finally $A(v)$ a remaining finite contribution
\cite{braaten:engl2}.
In Fig.~2 the result is summarized for a charm quark, where
for $\lambda_g=1, \lambda_q=0$ the loss is due
to elastic gluon-heavy quark scattering mediated by gluon exchange.

\begin{figure}
\begin{center}
   \epsfig{bbllx=94pt,bblly=264pt,bburx=483pt,bbury=587pt,
file=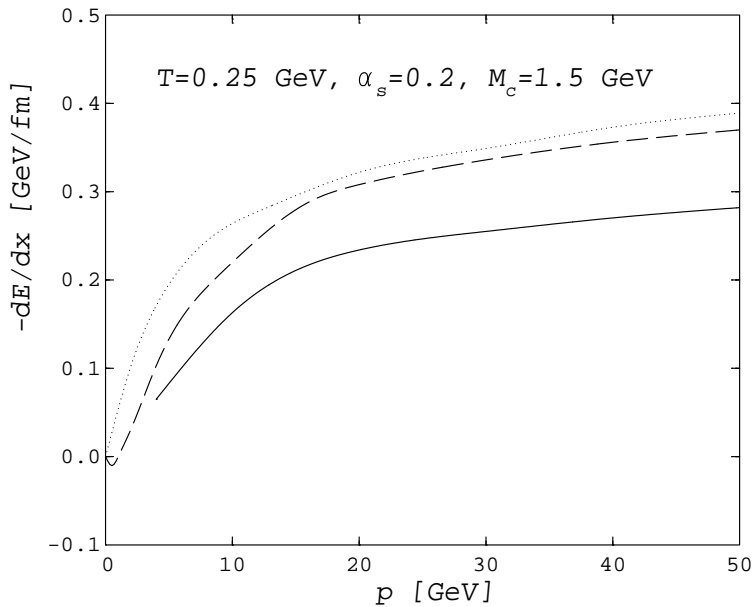, width=84mm,height=70mm}
\end{center}
\vskip 1truecm
\caption{Collisional energy loss
of a charm quark as a function of its momentum. The quark propagates through an
 out-of-chemical equilibrium plasma with fugacities
 $\lambda_g = 1, \lambda_q = 0$ (solid curve).
The dashed curve is the equilibrium result of \cite{braaten:engl2}, 
the  dotted curve shows the original prediction by Bjorken \cite{bjorken:engl}
.}
\label{fig2:fig}
\end{figure}

\section{Conclusion}
In this work we investigated interaction processes occurring in a 
plasma away from chemical equilibrium, namely the production of hard
real photons and the collisional energy-loss. The important observation
made is that dynamical screening is effective also away from equilibrium, and
finite predictions, (\ref{eq:resph}) and (\ref{eq:reseng}), can be given for 
the respective rates. Both rates depend on the fugacities 
$\lambda_q, \lambda_g$, so that specific predictions with respect to
actual heavy-ion experiments necessarily require further assumptions on 
the expansion and equilibration dynamics. 

Finally, it is necessary to point out that the absence of pinch singularities,
in the way observed here, may not be valid in general. Namely in the production
of leptons (virtual photons) contributions from time-like momentum exchange 
arise and a more elaborate approach might have to be developed
\cite{lebellac:pinch,niegawa:dilepton,niegawa:bolz}.

\section{Acknowledgments}
This work is supported in part by DFG project KA 1198/4-1. Thanks to D. Schiff
for stimulating discussions. 
M.D. likes to thank A. Ni\'egawa and the department of physics at Osaka City
University for kind hospitality as well as the German Academic Exchange 
Service (DAAD) for financial and general support of his stay in Osaka. 

\bibliographystyle{prsty}
\bibliography{pro}

\begin{thebibliography}{10}

\bibitem{htl1}
R.~D. Pisarski, Physica {\bf A158},  146  (1989).

\bibitem{htl2}
E. Braaten and R.~D. Pisarski, Nucl. Phys. {\bf B337},  569  (1990).

\bibitem{htl3}
J. Frenkel and J.~C. Taylor, Nucl. Phys. {\bf B334},  199  (1990).

\bibitem{kapusta:photon}
J. Kapusta, P. Lichard, and D. Seibert, Phys. Rev. {\bf D44},  2774  (1991).

\bibitem{baier:photon}
R. Baier, H. Nakkagawa, A. Ni{\'e}gawa, and K. Redlich, Z. Phys. {\bf C53},
  433  (1992).

\bibitem{shuryak:nephoton}
E. Shuryak and L. Xiong, Phys. Rev. Lett. {\bf 70},  2241  (1993).

\bibitem{strickland:nephoton}
M. Strickland, Phys. Lett. {\bf B331},  245  (1994).

\bibitem{kaempfer:nephoton}
B. K{\"a}mpfer and O. Pavlenko, Z. Phys. {\bf C62},  491  (1994).

\bibitem{thoma:nephoton}
M.~H. Thoma and C.~T. Traxler, Phys. Rev. {\bf C53},  1348  (1996).

\bibitem{muller:nephoton}
D.~K. Srivastava, M.~G. Mustafa, and B. M{\"u}ller, Phys. Rev. {\bf C56},  1064
   (1997).

\bibitem{bjorken:engl}
J.~D. Bjorken, Fermilab Report PUB-82/59-THY (unpublished)  (1982).

\bibitem{braaten:engl1}
E. Braaten and M.~H. Thoma, Phys. Rev. {\bf D44},  1298  (1991).

\bibitem{braaten:engl2}
E. Braaten and M.~H. Thoma, Phys. Rev. {\bf D44},  2625  (1991).

\bibitem{baier:nephoton}
R. Baier, M. Dirks, K. Redlich, and D. Schiff, Phys. Rev. {\bf D56},  2548
  (1997).

\bibitem{baier:neegloss}
R. Baier, M. Dirks, and K. Redlich, Talk given at 5th International Workshop on
  Thermal Field Theories and Their Applications, Regensburg, Germany, 10-14 Aug
  1998 (hep-ph/9809214)  (1998).

\bibitem{shuryak:hotglue}
E. Shuryak, Phys. Rev. Lett. {\bf 68},  3270  (1992).

\bibitem{biro:therma}
T.~S. Biro {\it et~al.}, Phys. Rev. {\bf C48},  1275  (1993).

\bibitem{kapusta:cascade}
K. Geiger and J.~I. Kapusta, Phys. Rev. {\bf D47},  4905  (1993).

\bibitem{wong:therma}
S.~M.~H. Wong, Nucl. Phys. {\bf A638},  527C  (1998).

\bibitem{kaempfer:nedilepton}
B. K{\"a}mpfer, O.~P. Pavlenko, A. Peshier, and G. Soff, Phys. Rev. {\bf C52},
  2704  (1995).

\bibitem{landsman:rev}
N.~P. Landsman and C.~G. van Weert, Phys. Rept. {\bf 145},  141  (1987).

\bibitem{chou:noneq}
K.-C. Chou, Z.-B. Su, B.-L. Hao, and L. Yu, Phys. Rept. {\bf 118},  1  (1985).

\bibitem{heinz:noneq}
S. Mrowczynski and U. Heinz, Ann. Phys. {\bf 229},  1  (1994).

\bibitem{calzetta:noneq}
E. Calzetta and B.~L. Hu, Phys. Rev. {\bf D37},  2878  (1988).

\bibitem{henning:noneq}
P.~A. Henning, Phys. Rept. {\bf 253},  235  (1995).

\bibitem{altherr:pinch1}
T. Altherr and D. Seibert, Phys. Lett. {\bf B333},  149  (1994).

\bibitem{altherr:pinch2}
T. Altherr, Phys. Lett. {\bf B341},  325  (1995).

\bibitem{braaten:screening}
E. Braaten and T.~C. Yuan, Phys. Rev. Lett. {\bf 66},  2183  (1991).

\bibitem{thoma:noneq1}
M.~E. Carrington, H. Defu, and M.~H. Thoma, Eur. Phys. J. {\bf C7},  347
  (1999).

\bibitem{lebellac:pinch}
M. {Le~Bellac} and H. Mabilat, Z. Phys. {\bf C75},  137  (1997).

\bibitem{niegawa:dilepton}
A. Ni{\'e}gawa, Eur. Phys. J. {\bf C5},  345  (1998).

\bibitem{niegawa:bolz}
A. Ni\'egawa, Prog. Theor. Phys. {\bf 102},  1  (1999).

\end{thebibliography}

\end{document}